# Adverse effect of ethanol on insulin dimer stability

Puja Banerjee, Sayantan Mondal and Biman Bagchi*

Solid State and Structural Chemistry Unit

Indian Institute of Science, Bangalore-560012, India

* Corresponding Author's e-mail: *bbagchi@iisc.ac.in*

___________________________________________________________________

## Abstract

*Alcohol is widely believed to have an effect on diabetes, often considered beneficial in small amounts but detrimental in excess. The reasons are not fully known but questions have been asked about the stability of insulin oligomers in the presence of ethanol. We compute the free energy surface (FES) and the pathway of insulin dimer dissociation in water and in 5% and 10% water-ethanol mixture. We find that in the presence of ethanol the barrier energy of dissociation reaction decreases by about 40% even in 5% water-ethanol solution. In addition, ethanol induces a significant change in the reaction pathway. We obtain estimates of the rate of reaction and binding energy for all the three systems and those agree well with the previous experimental results for the insulin dimer dissociation in water. The computed FES in water exhibits ruggedness due to the existence of a number of intermediate states surrounded by high and broad transition state region. However, the presence of ethanol smoothens out the ruggedness. We extracted the conformations of the intermediate states along the minimum energy pathway in all the three systems and analyzed the change in microscopic structures in the presence of ethanol. Interestingly, we discover a stable intermediate state in water-ethanol mixtures where the monomers are separated (center-to-center) by about 3 nm and the contact order parameter is close to zero. This intermediate is stabilized by the distribution of ethanol and water molecules at the interface and which, significantly, serves to reduce the dissociation rate constant .The solvation of the two monomers during the dissociation and proteins' departure from native state configuration are analyzed to obtain insight into the dimer dissociation processes.*

___________________________________________________________________

**Table of contents graphic:**

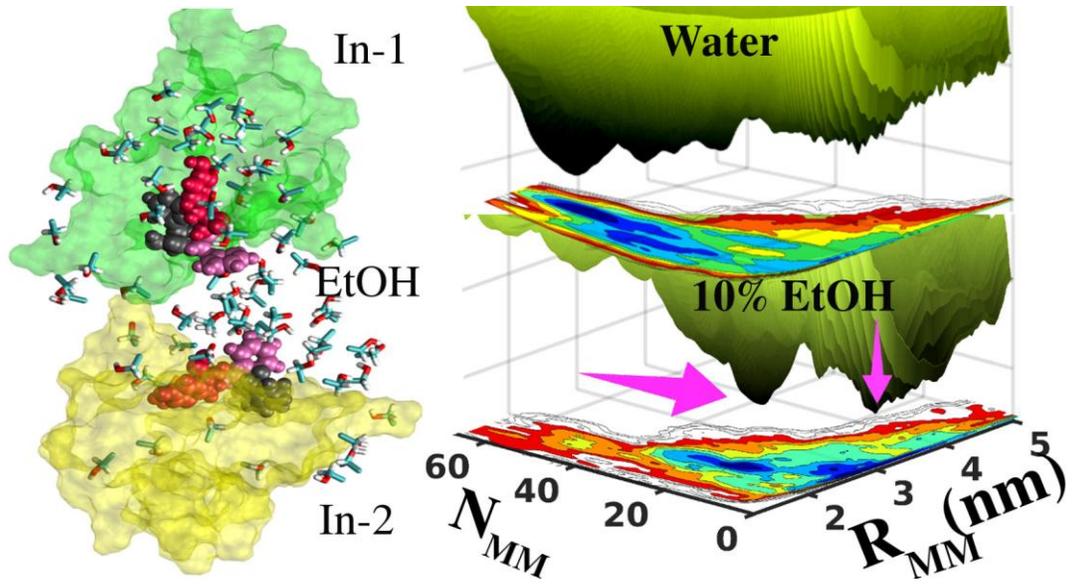

# I. Introduction:

Protein association and dissociation are general problems of intense current interest[1,2,3,4,5,6]. Of late, dissociation-association of protein insulin have drawn particular attention because of their role in the control of human blood sugar. While monomer is the biologically active species, hexamer serves as a storage in human body while dimer is responsible for endogenous delivery. Therefore, dissociation of dimer into monomer, and the reverse process of association are important biochemical processes that have been studied recently by sophisticated experimental methods like 2D-IR spectroscopy, as discussed below.

Effects of ethanol on human body functions have long been a debated issue. An increasing amount of attention has been devoted to the effects of ethanol on diabetes. Many previous studies have demonstrated that ethanol consumption could act as a risk factor to develop type II diabetes mellitus at a heavy consumption while beneficial in case of moderate consumption[7,8,9]. It was also pointed out that alcohol affects glucose and insulin metabolism in a number of ways. Naturally, the focus has centered upon the effects of alcohol on the physiological role of insulin.

A number of previous experimental and computational studies have interrogated the stability, dynamics and mechanism of dissociation of the insulin dimer [5,6,10,11,12,13,14,15,16,17,18,19]. Tokmakoff *et al.* studied dimer dissociation of insulin employing 2D-IR spectroscopy by probing amide-I mode as its $\nu_\perp$ mode is delocalized over two monomers in stable dimeric state and vanishes upon dissociation[5,6]. They have calculated the dissociation constant of dimer in water and also observed that the presence of ethanol in the medium increases the dissociation constant. These authors have also investigated the dynamics of insulin dimer dissociation using laser induced temperature jump (T-jump) experiments and observed the formation of insulin dimer in 5-150 μs time scale[6].

Recently, Rimmerman *et al.* studied insulin association dynamics using time-resolved X-ray scattering in combination with ultrafast T-jump and detected five transient species i.e., intermediate states and the dynamics of them[19]. They have reported that intermediate states form much faster than 5 μs.

A molecular level understanding of these important experimental studies does not exist at present, despite the importance of insulin dimer dissociation process. In fact, there are very few studies on the dissociation of protein dimers. This is in contrast to the studies on protein association where many theoretical studies have investigated the entire reaction process [4,20,21,22,23]. We note here that association and dissociation are physically two quite different processes, with the free energy landscape playing a critical role in the latter process.

There exist a number of studies that addressed the binding energy by using molecular mechanics-generalized Born surface area (MM-GBSA) approach, pioneered by Karplus group[24]. In a well-known study, Karplus and coworkers calculated binding free energy change of insulin dimerisation to be -11.2 kcal/mol [24] which corroborates well with the experimental value of -7.2 kcal/mol[25]. Kim *et al.* carried out AFM experiment and steered MD simulation to show the relative strength of the inter-monomer interactions across the antiparallel *β*-sheet interface and intra-monomer interactions of A chain residues with B-chain plays critical roles in determining the conformational changes during dissociation [26].

However, there are no studies available that looked into the free energy surface as well as the microscopic details of insulin dimer dissociation. In this article we present a detailed molecular analysis of dimer dissociation in water and subsequently studied the changes in the energetics ,

reaction pathway and the rate of the reaction that are induced by the presence of ethanol in the medium.

We would like to point out that protein association is an active area of research almost for more than last half-century for multiple biological and medical reasons that have far reaching consequences on human health [27,28,5,3,29]. One of the most studied cases perhaps is the association of β-amyloid that is related to the Alzheimer disease[30,31]. It is believed that some proteins tend to associate in the non-native or unfolded state. Several osmolytes, for example, urea, DMSO, ethanol etc., are known to promote unfolding of protein, and therefore can lead to protein aggregation. However, the reverse scenario is also shown to be true for a different set of osmolytes. Sometimes these processes are dependent on the concentration of the co-solvents/co-solutes. Our earlier work reveals the effect on protein folding dynamics exerted by different concentration of water-ethanol binary mixtures[32].

Given the importance of insulin in the physiological function of our body, the lack of understanding of the stability and activity of insulin oligomers in water and in presence of various solutes is surprising. We note that while protein association is often believed to cause serious ailments, in the case of insulin the reverse could be true in the sense that dimer and hexamer serve important positive biological functions as storage of the monomer. Earlier we have focused on the structural and dynamical features of hexamer, dimer or monomeric inulin in aqueous solution [33,34]. Recently, Singh et al. have analysed folding of insulin monomer emplying bias-exchange metadynamics technique[35]. Like many other proteins, insulin forms oligomers by modifying its native folds. Since monomer is the biologically active form, dissociation of dimer and hexamer is subject of intense interest.

Insulin is a 51 residue dual-chain hormone. In the monomeric form, insulin consists of two chains, A and B, which contain 21 and 30 residues respectively. In human body, following the synthesis of monomeric insulin, two monomers associate into a dimer by the formation of four hydrogen bonds which leads to the formation of anti-parallel intermolecular β sheets at the C-termini of chain B (residues: Phe-24, Phe-25 and Tyr-26) as shown in **Figure 1**.

A number of previous studies were devoted to analyze the structural features of insulin dimer. C-termini end of the B chain (B22-B30) of insulin monomer has been shown to stabilize the dimer through the formation of antiparallel β-sheet[36, 37] (**Figure 1**). Residues B24-B26 (PHE and TYR) are involved in hydrophobic interactions and in the formation of intermolecular bridging hydrogen bonds. Recent molecular dynamics simulation reveals that the structural stability arises because of non-bonded correlated molecular motions[38]. Self-association can be controlled by incorporating site-specific modifications to develop bioactive insulin analogue in monomeric form.

Three of the dimers further self-assemble by coordinating with two bivalent Zn ions through six His-10 residues and stored in the zinc rich vesicles in the β-cells of pancreas as hexamers. However, the active form of insulin hormone is the monomeric form that interacts with the receptor located on the cell membrane[39]. In Eq.(1) we provide a simple and minimalistic illustration of the above mentioned process.

$$\begin{aligned}\left[(In)_6 \cdot 2Zn^{2+}\right](aq) &\leftrightarrow 3(In)_2(aq) + 2Zn^{2+} \\ (In)_2(aq) &\leftrightarrow 2In(aq)\end{aligned} \qquad (1)$$

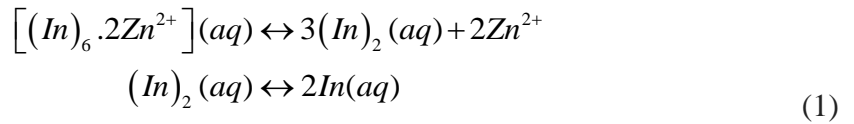

Here, we aim to primarily focus on the dimer dissociation process. At a constant pressure and thermodynamic equilibrium, this process is associated with an equilibrium constant for dimer dissociation,

$$K_{eq} = \frac{\left[(In)\right]^2}{\left[(In)_2\right]} \quad (2)$$

where $\left[(In)_2\right]$ and $\left[(In)\right]$ are the equilibrium concentration of dimer and monomer respectively. This quantity is most relevant in experimental studies like 2D-IR.

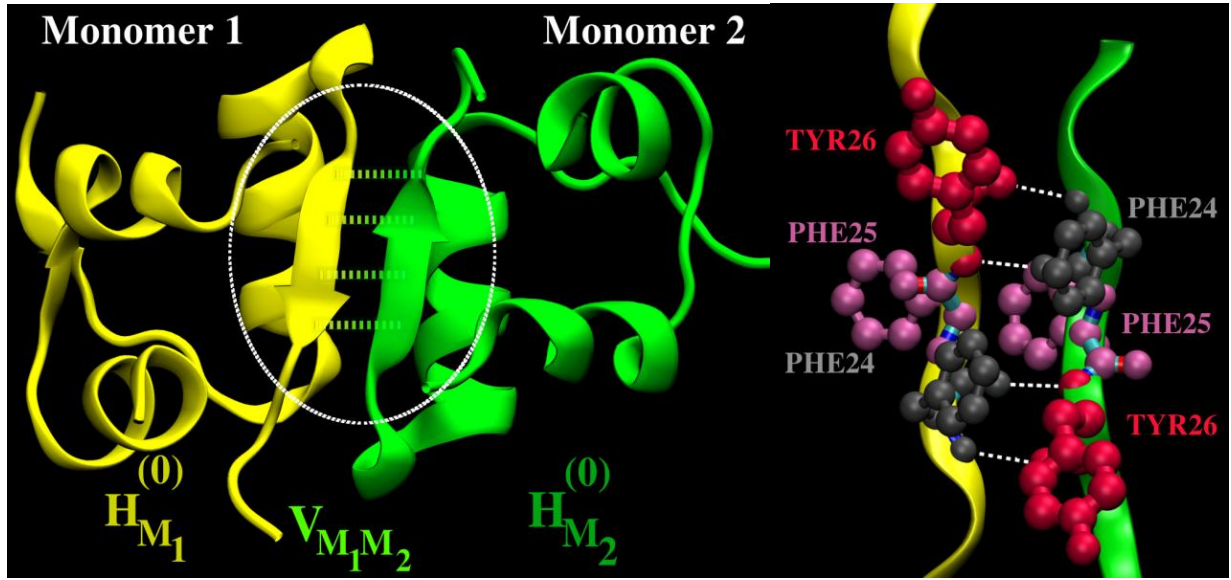

*Figure 1: Insulin dimer with the stabilizing intermolecular anti-parallel β sheet. Residues of B chain (Phe-B24, Phe-B25 and Tyr-B26) help in the formation of the antiparallel β-sheet by four hydrogen bonds through backbone atoms. Additional stability arises due to hydrophobic interactions amongst the phenyl alanine side-chains. The total Hamiltonian of the system seeks contribution from three terms: two pure Hamiltonians of individual monomers in the absence of the other monomer, $H_{M_1}^{(0)}$ and $H_{M_2}^{(0)}$ and the perturbation term due to the monomer-monomer interaction, $V_{M_1M_2}$.*

As shown in **Figure 1**, the total Hamiltonian of the dimer system can be expressed as the sum of individual monomeric interaction energies and the cross interaction between the atoms of monomer ($M_1$) and monomer-2 ($M_2$).

$$H_D = H_{M_1}^{(0)} + H_{M_2}^{(0)} + V_{M_1 M_2} \tag{3}$$

Now, let us define a potential energy of dimer with respect to two relevant collective variables, X and Y as $U_D(X,Y)$. Similarly two monomers have the potential of mean force (PMF) $U_{M_1}(X,Y)$ and $U_{M_2}(X,Y)$.

$$U_D(X,Y) = U_{M_1}(X,Y) + U_{M_2}(X,Y) + U_{M_1-M_2}^{cross}(X,Y) \tag{4}$$

Therefore, $U_{M_1-M_2}^{cross}(X,Y)$ contains information about the stability and also, partly, the dimer dissociation rate. In simulation we can calculate free energy of the dimeric system as a function of two order parameters. Here X and Y order parameters can be suitably chosen as the distance ($R_{MM}$) between the centre of masses of two monomers and the number of contacts between two monomers ($N_{MM}$). Then the dimer dissociation energy can be determined from the difference of free energy between dimeric and monomeric states.

In the present study, the simulation of dimer dissociation of two insulin monomers has been performed using parallel tempering metadynamics (PTMetaD) scheme in well-tempered ensemble (WTE), the technique introduced by Parinello and coworkers[40], as discussed in section II. For insulin monomer with 51 amino acid residues, the free energy calculation of dimer dissociation using general well-tempered metadynamics (WT-MetaD) is not sufficient because of its slow convergence. Therefore, we have used PTMetaD-WTE that helps us to attain a faster convergence and to reduce the computational cost of protein dimer dissociation simulation with all-atom model for protein and explicit solvent molecules.

First we have studied the dissociation process of insulin dimer in water by investigating the free energy surface (FES) with respect to two collective variables: (i) center-of-mass distance between two monomers ($R_{MM}$) and (ii) number of contacts between the monomers ($N_{MM}$). Then we have modified the solvent into water-ethanol binary mixture with ethanol mole fraction 5% and 10% and obtain the free energy surface (FES) with respect to the same set of collective variables (section III.A.). In section III.B., we have discussed the free energy change along minimum energy pathway for all the three systems with different solvent and estimated the rate of dissociation. *Computed binding free energy of insulin dimer in water agrees well with the previous experimental study of Strazza et al. and the simulations of Karplus et al.* [10,13]. The results are also in general agreement with the conclusions arrived at by Tokmakoff *et al.* by 2D-IR experiments coupled with temperature jump (T-jump)[6] and Rimmerman *et al.* by time-resolved X-ray scattering experiments coupled with ultrafast T-jump[19].

In section III.C., we have presented a detailed analysis of the microscopic structure along the reaction pathway with the intermediate state structures. We have also focused on the structural changes induced because of ethanol, by computing the relative orientation and distance of residues in the antiparallel β-sheet region and distance matrices of dimeric cross-contacts. Finally, we have analyzed the solvation of the intermediate states in section III.D. and connected the results of solvation with microscopic structural changes in order to explain the change in free energy surface, the rate of dissociation reaction and reaction pathway (with different intermediate state structures) in the presence of ethanol.

## II. Computational Methods

We have used GROMACS-5.0.7[41] for molecular dynamics simulations and PLUMED-2.0 plugin[42] patched to GROMACS for metadynamics calculations. CHARMM36 force-field parameters[43] have been used for proteins, TIP3P model for water molecules[44] and ethanol molecules have been modeled using generalized CHARMM force-field (CGenFF)[45, 46]. The details about the size of the system simulated are given in the subsequent sections.

### A. Molecular Dynamics Simulations

We have taken the initial structure for molecular dynamics simulation from the X-ray structure of insulin dimer resolved at 0.92 Å (PDB ID: 3W7Y). We have added four $Na^+$ ions in the simulation box to maintain electrostatic charge neutrality. Energy minimization of the system has been performed using the steepest descent algorithm and the system has been equilibrated in NPT ensemble (T=300K and P=1 bar) for 2 ns and then in NVT ensemble (T=300K) for 20ns.

All covalent bonds have been constrained to their respective equilibrium values by LINCS algorithm[47]. Both the Lennard-Jones and Coulomb interactions have been calculated using a cutoff of 0.9 nm. Long range electrostatic interactions have been calculated using Particle Mesh Ewald (PME) algorithm[48].

Next, we have solvated the insulin dimer with binary mixtures of water and ethanol containing 5% and 10% mole fraction of ethanol. Here, we have followed similar system size and simulation details as discussed above.

To characterize different intermediate states along the minimum energy pathway (MEP) of free energy surface in all three systems, we have extracted the most probable configuration in the particular well from metadynamics simulation and an equilibrium simulation has then been carried out. We have equilibrated the system in NPT ensemble (T=290K and P=1bar) for 2 ns. Subsequently, a trajectory of 10 ns has been produced in NVT ensemble, dumping data at every 0.02 ps interval for analysis.

In addition, we have carried out separate sets of simulations with the monomeric insulin in water-ethanol binary mixtures as stated above. The parameters and simulation details of these simulations have been taken the same as discussed before.

## B. Metadynamics

We have introduced metadynamics bias on the equilibrated structures following parallel tempering metadynamics in well-tempered ensemble (PTMetaD-WTE) method with twelve replicas distributed in the temperature range 290-620 K. At first, we have applied bias potential in the energy space for 20 ns with. This helps the actual metadynamics simulation with different order parameters to converge faster by increasing the fluctuation in energy space.

In this study, we have considered two order parameters (or collective variables, CVs) to study the dimer dissociation pathway. The order parameters chosen are (i) the distance between the center-of-mass of two monomeric units ($R_{MM}$), and (ii) the number of cross contacts between the $C_\alpha$ atoms of two monomers ($N_{MM}$). Then, we have carried out PTMetaD in well-tempered ensemble by restarting metadynamics simulation after the previous 20ns metadynamics run in energy space. We have checked the convergence of the metadynamics simulations by examining the free energy profile at certain intervals.

## III. Results and Discussions

## A. Free energy surface of insulin dimer dissociation in water and water-ethanol binary mixtures

**Figure 2** shows the free energy surface (FES) contour of the insulin dimer dissociation process in water as a function of two collective variables (CVs): the distance between the center-of-mass (COM) of two monomers ($R_{MM}$) and the number of $C_\alpha$-$C_\alpha$ contacts between the two monomers ($N_{MM}$). Regardless of the ruggedness, the FES clearly shows a global minimum that corresponds to the dimer ($R_{MM}$~1.8 nm and $N_{MM}$ ~ 50, indicated by 'A' in *Figure 2*). There are many local minima with a broad transition state and a high barrier region ($R_{MM}$~2.4 nm and $N_{MM}$ ~ 15). However, we have computed the minimum energy path and indicated some local minima on this path by 'B', 'C' and 'D' with higher energy than conformation 'A'(Relative free energy values are written in **Figure 2**). State 'B' corresponds to a lower monomeric distance ($R_{MM}$~1.7 nm) but with a lower number of contacts ($N_{MM}$~32). Upto State C ($R_{MM}$~1.9 nm, $N_{MM}$~20), there is no significant change in intermonomeric distance, but number of contacts decreases to a large extent. The interesting point to note here is, the minimum energy path depicts that the number of contacts ($N_{MM}$) changes significantly upto ~10 during dissociation of dimer without any significant change in the COM distance ($R_{MM}$). This path clearly avoids the higher barrier region as well as some local minima due to higher free energy barriers surrounding those states.

We have shown the most probable representative conformations corresponding to four minima in the same figure. Microscopic structural changes in the dimer forming region with antiparallel β-sheet as well as the protein conformational changes act as a driving factor in the dissociation reaction. We shall discuss the structural changes in addition to the energetics of the minimum energy path later in detail.

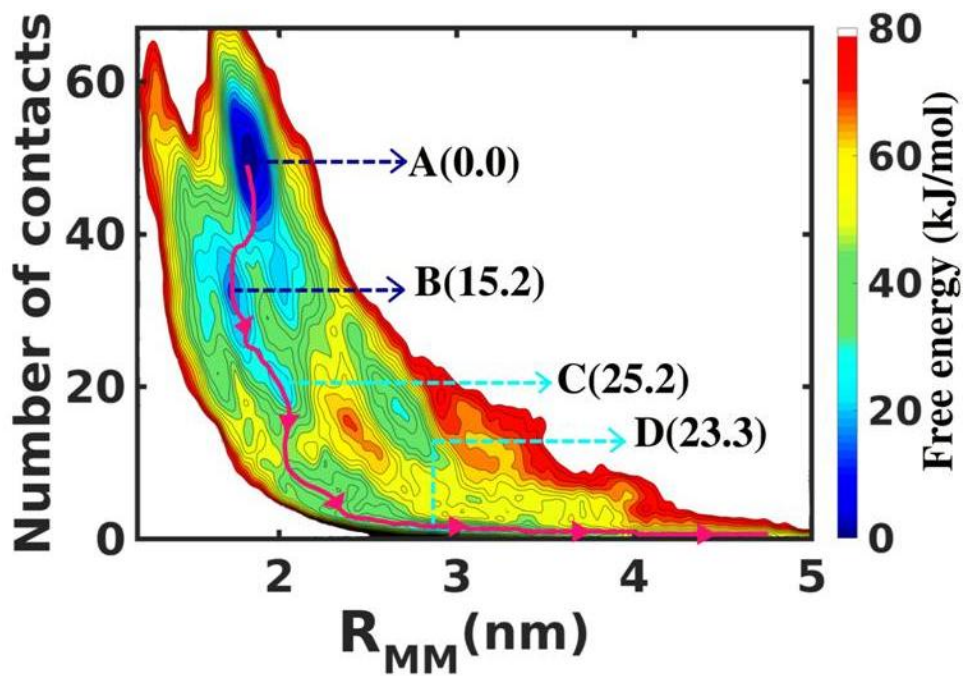

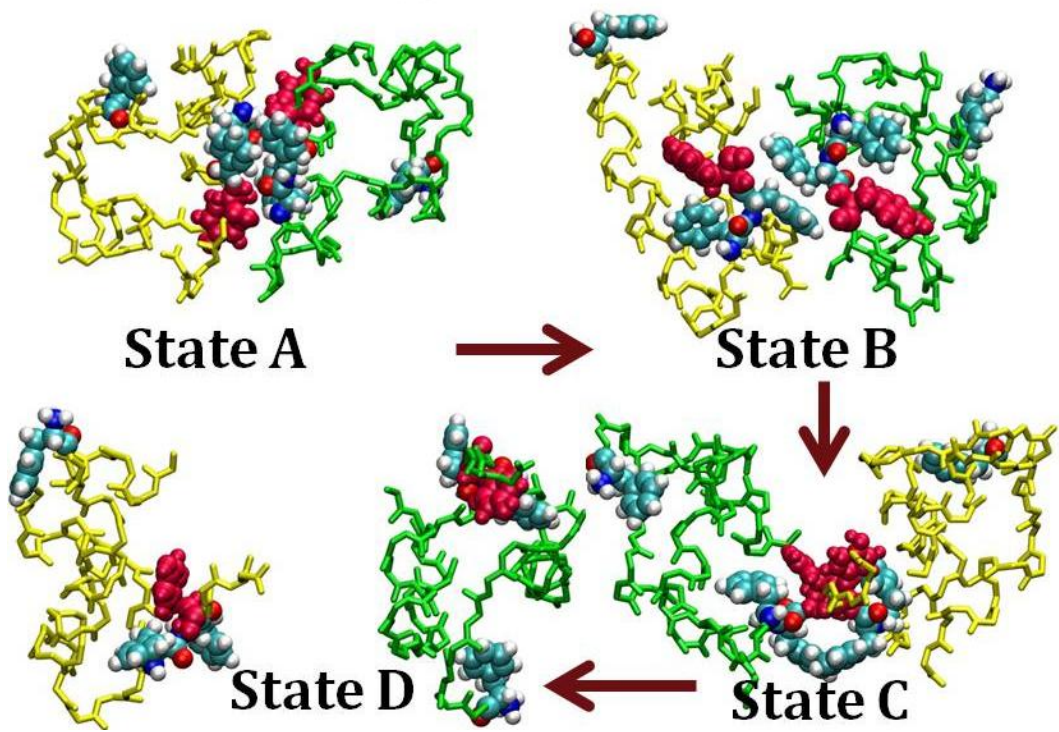

*Figure 2: Free energy surface(FES) of insulin dimer dissociation in water as a function of center-of-mass distance of two monomers ($R_{MM}$) and number of contacts ($N_{MM}$). Red color in*

*FES indicates the highest energy regions such as the barrier region, separated monomeric region ($R_{MM}$~5 nm). The most stable dimeric state corresponding to global minima with some local minima are pointed out in the FES (with the free energy values in kJ/mol in the brackets) and the representative configurations are also shown. Minimum energy path from state A to separated monomeric state is shown (pink line).*

Now, we aim to study the dissociation pathway of the insulin dimer in the presence of a co-solvent (here ethanol) and how it might affect the energetics of the dimer dissociation. To achieve this goal, we have carried out the similar advanced metadynamics scheme to obtain the FES of dimer dissociation in presence of ethanol. Here we have taken two systems: 5% and 10% mole fraction of aqueous ethanol.

**Figure 3** shows the FES of dimer dissociation in presence of 5% ethanol in the medium. Compared to the FES in water, here the global minima (denoted as state 'A') corresponds to the stable dimeric state is found to have less number of contacts ($N_{MM}$~32) with a similar inter-monomeric distance ($R_{MM}$~1.89 nm). Similar to the previous one, here also we have computed minimum energy path (MEP) (pink line shown in **Figure 3**) connecting the most stable dimeric state, 'A' and the separated monomeric state ($R_{MM}$~5 nm, $N_{MM}$=0). The most important point to note here is that not only the structure of the dimeric state and the energy barrier change in presence of ethanol, but there is a significant change in the pathway of the reaction. Unlike in water, here the MEP involves a parallel change in both the order parameters $R_{MM}$ and $N_{MM}$ starting from state 'A' upto state 'C' untill the number of contacts ($N_{MM}$) becomes zero.

The most probable conformations of the protein in states 'A', 'B' and 'C' are shown in **Figure 3**. Being an amphiphilic solvent, ethanol offers both hydrophobic interactions to the hydrophobic amino acid residues, at the same time interacts with polar amino acids and induces significant conformational changes in the protein. Therefore, the stable dimeric form as well as intermediate states have considerably different structures as well as energetics compared to the system in pure water which leads to a change in reaction pathway as well as dissociation energy and rate of the dissociation reaction.

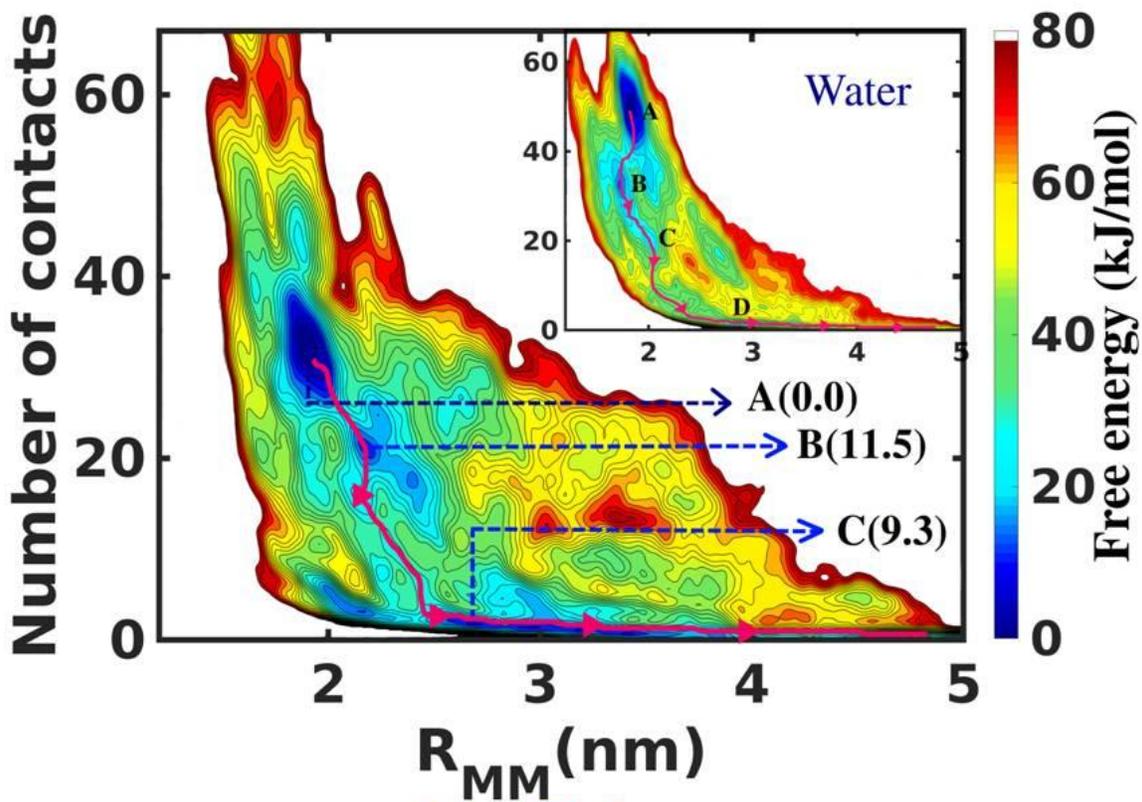

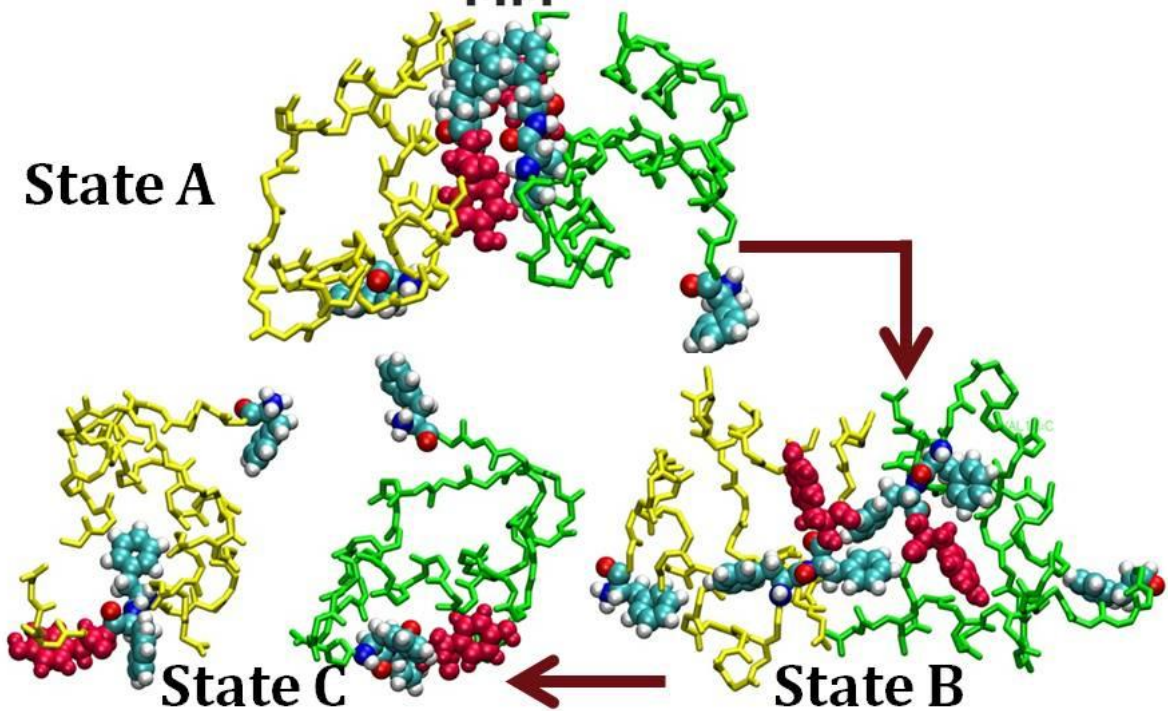

*Figure 3: Free energy surface(FES) of insulin dimer dissociation in 5% water-ethanol binary mixture as a function of center-of-mass distance of two monomers ($R_{MM}$) and number of contacts ($N_{MM}$). Red color in FES indicates the highest energy regions such as the barrier region, separated monomeric region ($R_{MM}$~5 nm) etc. The most stable dimeric state corresponding to global minima with some local minima are pointed out in the FES and the representative configurations are also shown. Free energy values (in kJ/mol) corresponding to each state are reported in brackets. For comparison the FES in pure water is shown in inset.*

The interaction of ethanol with the protein residues leads to a much more pronounced deviations from the native dimer structure in 10% ethanol-water mixture. **Figure 4** shows that the free energy minimum corresponds to the dimer in this ethanol concentration shifts to a completely different configuration with a large intermonomeric distance ($R_{MM}$~2.5 nm) and small number of contacts between two monomers ($N_{MM}$~18). We have shown the minimum energy path (pink line) and some intermediate states on this path. Surprizingly, the intermediate state, C is the global minima of this FES where number of contacts ($N_{MM}$) reduces to ~1 and the intermonomeric distance ($R_{MM}$) increases to ~3.1 nm.

We have shown the representative structures of the three states in **Figure 4**. They exhibit considerably different protein conformations where state A and B correspond to more extended protein chains, in state C, protein chains are a bit contracted. We have already discussed that ethanol participates in the hydrophobic interaction with the hydrophobic amino acid residues such as PHE, ILE, VAL etc. In state C, system attains minimum energy by the compensation of entropy by the enthalpy where the ethanol molecules in the medium can have a favorable interaction with the PHE residues in the junction of two monomers as they are considerably separated.

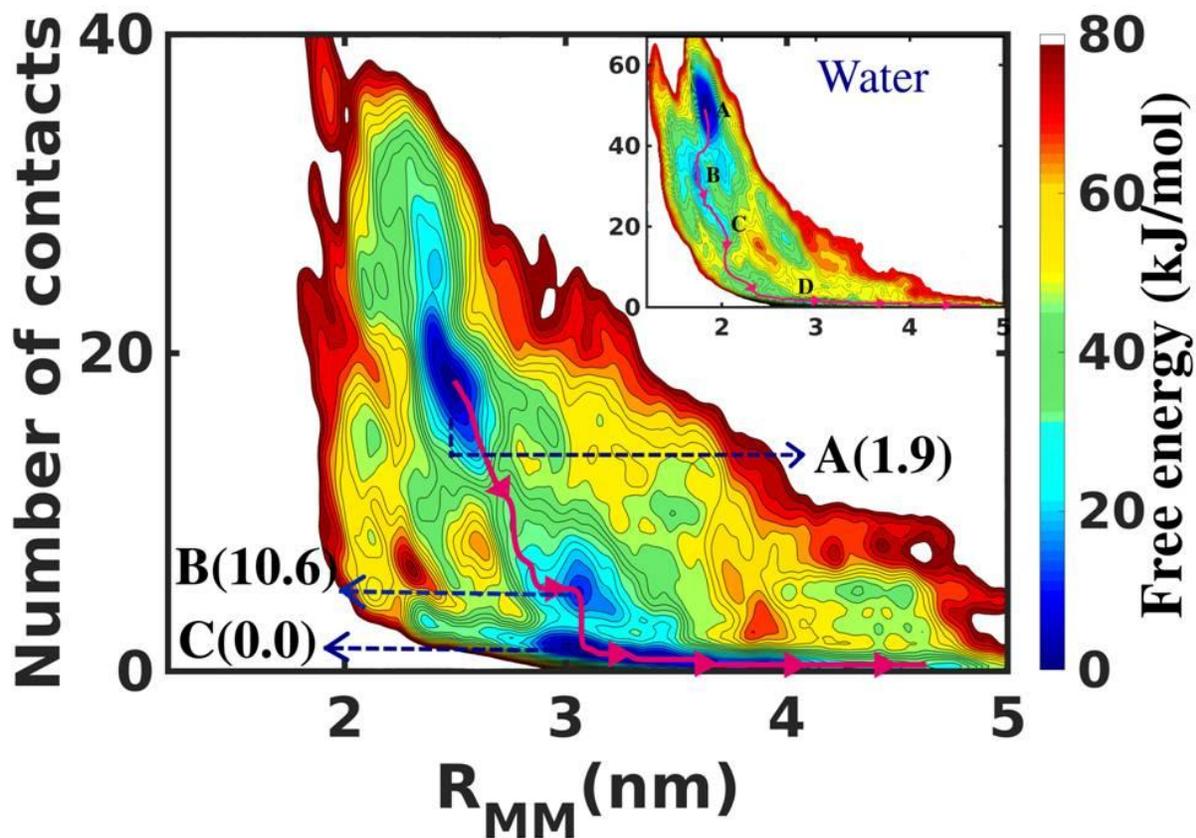

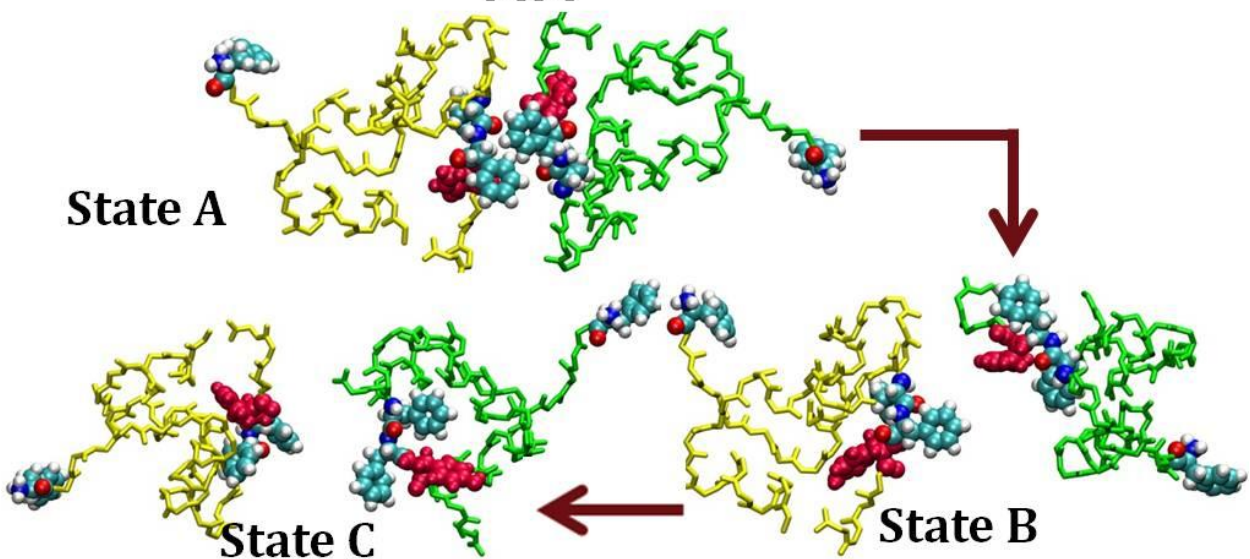

*Figure 4: Free energy surface(FES) of insulin dimer dissociation in 10% water-ethanol binary mixture as a function of center-of-mass distance of two monomers ($R_{MM}$) and number of contacts ($N_{MM}$). Red color in FES indicates the highest energy regions such as the barrier region, separated monomeric region ($R_{MM}$>5 nm). The most stable dimeric state*

*corresponding to global minima with some local minima are pointed out in the FES and the representative configurations are also shown. Free energy values (in kJ/mol) corresponding to each state are reported in brackets. For comparison the FES in pure water is shown in inset.*

We have compared the evolution of the two CVs in different intermediate states along the minimum energy path in pure water and in presence of ethanol in the system (**Figure 5**). The stable dimeric form (state 'A') in both pure water and in 5% water-ethanol mixture correspond to similar inter-monomeric distance ($R_{MM}$), compared to these, $R_{MM}$ of state 'A' in case of 10% water-ethanol mixture is significantly higher. However, number of contacts between two monomers ($N_{MM}$) for state 'A' shows a monotonic decrease from pure water to 5% ethanol to 10% ethanol due to protein conformational change. These data suggest the dramatic change in the configuration of the stable form of dimer in presence of ethanol in the system.

However, **Figure 5** clearly depicts that not only the conformation of the dimeric state changes but also the presence of ethanol affects the reaction pathway. In pure water, upto state 'C', $R_{MM}$ does not change significantly though $N_{MM}$ decreases to a large extent. But in 5% ethanol, there is a sharp increase in state 'B' and 'C'. The most dramatic change of reaction path occurs in 10% ethanol where starting from state 'A' during dissociation process it shows a sharp increase in $R_{MM}$ and decrease in $N_{MM}$ in all the intermediate states of dissociation process.

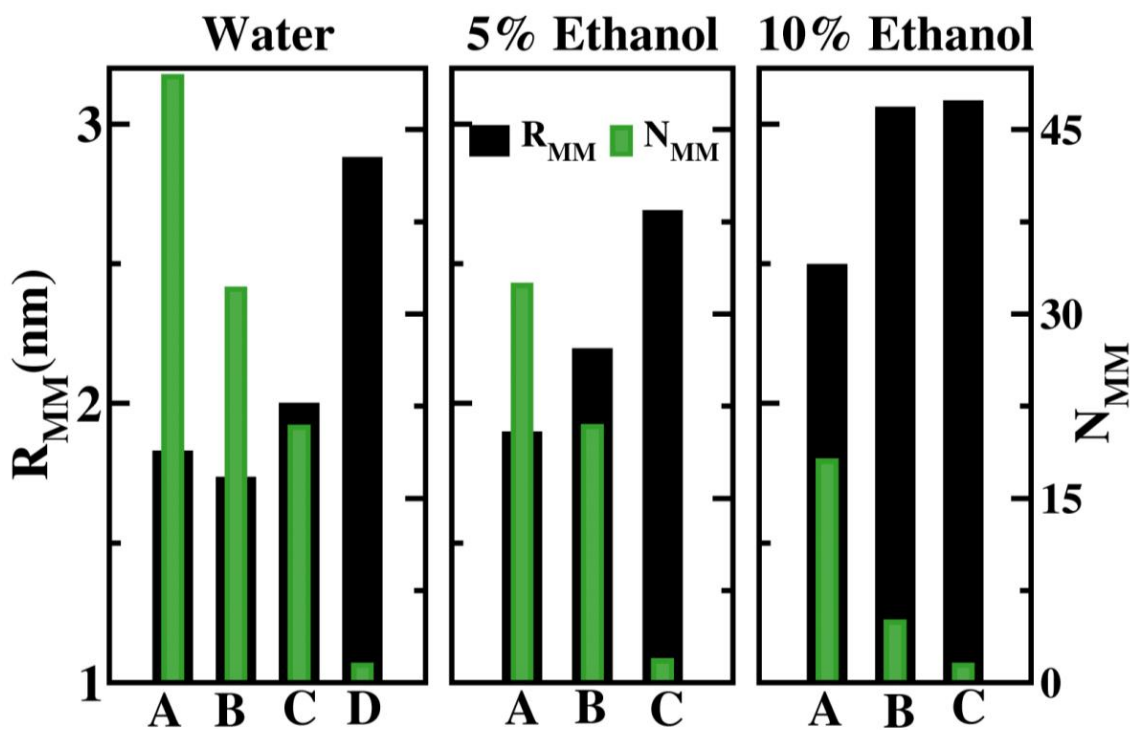

*Figure 5: Change in the two order parameters, inter-monomeric distance ($R_{MM}$) and number of contacts between two monomers ($N_{MM}$) in the different intermediate states of the free energy surface of dimer dissociation of insulin in pure water, 5% ethanol and 10% ethanol medium. This depicts the change in reaction pathway in presence of ethanol.*

### B. Effect of ethanol on the free energy change along the reaction pathway

Here we report the free energy of insulin dimer dissociation as a function of two CVs: $R_{MM}$ and $N_{MM}$ along the minimum energy pathway in three different systems of pure water and water-ethanol binary mixtures in **Figure 6**. In water, the FES involves significant ruggedness, however, ethanol in the medium smoothens out the ruggedness of FES. The binding energy of the insulin dimer is obtained as 51.9 kJ/mol where in the 5% and 10% ethanol medium it is reduced to 34 kJ/mol and 28.2 kJ/mol respectively. The largest barrier in the minimum energy pathway is observed as ~50 kJ/mol in water, ~30 kJ/mol in 5% ethanol and ~40 kJ/mol in 10% ethanol solution. In water, not only the first barrier around the stable dimer (state 'A') is considerably higher, but due to the higher binding energy and rugged landscape with multiple minima, overall free energy cost is much higher than the situation in presence of ethanol.

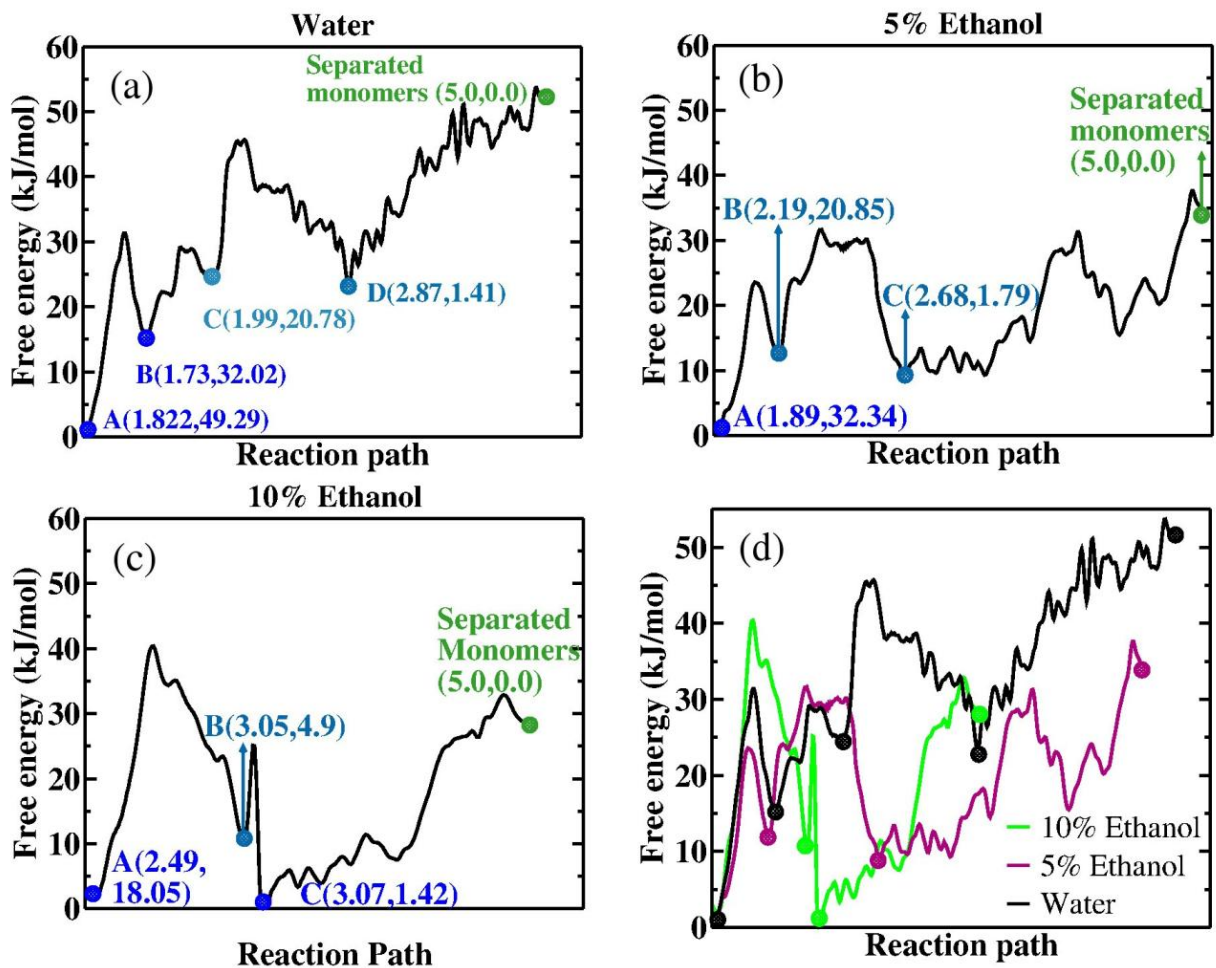

*Figure 6: Free energy of dimer dissociation along the minimum energy path (connecting stable dimeric form and separated monomeric form) in a system of: (a) pure water, (b) 5% water-ethanol mixture, (c) 10% water-ethanol mixture, (d) comparison of the free energy cost in the three systems. This is the two-dimensional free energy as a function of two order parameters: $R_{MM}$ and $N_{MM}$ the values of which corresponds to each states are written in the brackets.*

The calculation of the rate of dissociation of insulin dimer in water and in the presence of ethanol from this multidimensional free energy surface is prohibitively difficult. But an estimate of the rate can be given by extracting the free energy data in one dimension for both the CVs (**Figure 7**).

According to Eyring equation of transition state theory (with transmission rate coefficient=1),

$$k^{TST} = \frac{k_B T}{h} \exp\left(-\frac{\Delta G^{\ddagger}}{k_B T}\right) \quad (5)$$

Here $\Delta G^{\ddagger}$ is the free energy difference between the reactant and the activated transition state. Using the $\Delta G^{\ddagger}$ values from the one dimension free energy functions, $\Delta G(R_{MM})$ and $\Delta G(N_{MM})$ we obtain the rate of reaction that are tabulated in **Table 1**. The rate of dimer dissociation in water using $\Delta G^{\ddagger}(R_{MM})$ is obtained as 0.58 μs$^{-1}$ that matches well with the previous experimental results[6][19].

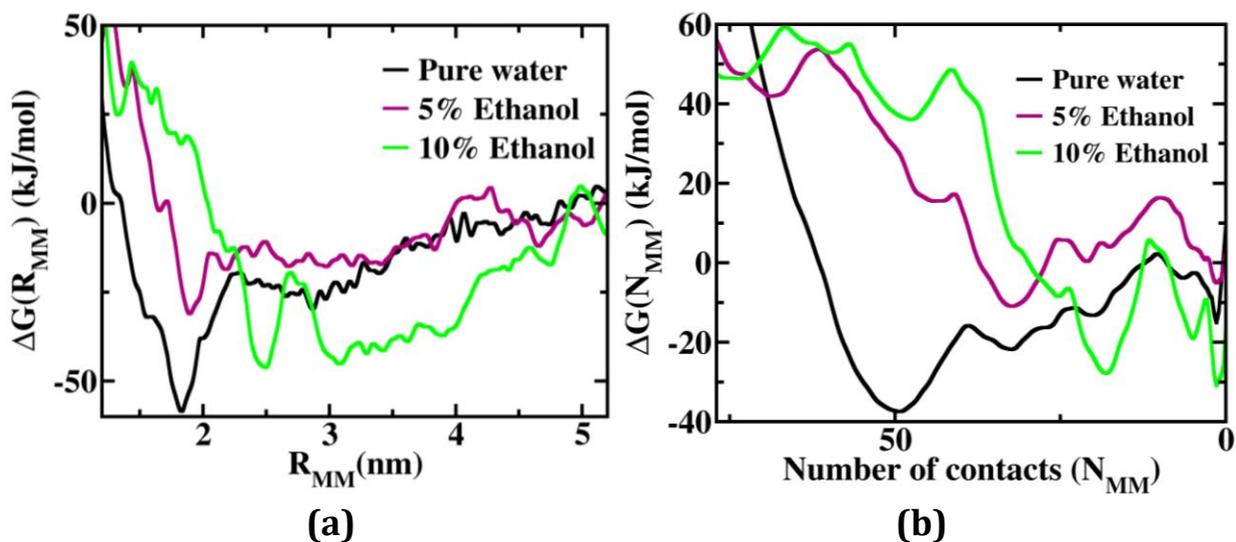

*Figure 7: One dimensional free energy as a function of the order parameters (a) inter-monomeric distance ($R_{MM}$), (b) number of contacts ($N_{MM}$) (defined earlier).*

*Table 1: Free energy barrier of the reactant dimeric species, $\Delta G^{\ddagger}$ in three different medium and the rate of reaction, $k^{TST}$ calculated from one dimensional free energy functions using Eyring equation of transition state theory.*

| 1D Free energy | Medium of reaction | $\Delta G^{\ddagger}$ (kJ/mol) | Rate constant, $k^{TST}$ |
|---|---|---|---|
| $\Delta G(R_{MM})$ | Water | 38.96 | 0.58 μs$^{-1}$ |
| | 5% water-ethanol | 17.11 | 5.0 ns$^{-1}$ |
| | 10% water-ethanol | 26.65 | 0.1 ns$^{-1}$ |
| $\Delta G(N_{MM})$ | Water | 21.61 | 0.77 ns$^{-1}$ |
| | 5% water-ethanol | 16.81 | 5.7 ns$^{-1}$ |
| | 10% water-ethanol | 33.56 | 5.4 μs$^{-1}$ |

## C. Microscopic structural changes of antiparallel β-sheet vs protein conformational changes:

Till now, we have discussed the energetics of dimer dissociation and the rate of reaction in water and in the presence of ethanol in the medium. Now, our aim is to investigate the dimer dissociation process at molecular level in these three systems to unravel how the solvent medium affects microscopic structural rearrangement that drives the reaction towards two separated monomers in different pathway and in different rate.

At first, we have analyzed the conformational changes of the two monomers in the junction, mainly the antiparallel β-sheet region. To this aim, we have plotted the dihedral angle between the backbone Cα in the antiparallel β-sheet region and the distances between each of those residues (PHE24, PHE25, TYR26) that forms the β-sheet. The dihedral angle is defined by the four backbone Cα of PHE-24(monomer 1), TYR-26(monomer 1), TYR-26 (monomer 2), PHE-24(monomer 2).

Thorough the change of the dihedral angle, **Figure 8** shows that the movement of the backbone in the antiparallel β-sheet region during the dimer dissociation for water (upto state 'C') as well as two water-ethanol mixture is smaller compared to the movement of the side-chains of the amino acid residues. Hence, the weak hydrophobic interaction gets compromised to retain the stabilizing intermolecular hydrogen bonding in the antiparallel β-sheet region by the enthalpy-entropy compensation.

Besides this, the important point to note here is, in the two ethanol-water mixtures, the change in the dihedral angle with the residue-residue distance in the antiparallel β-sheet region does not deviate significantly compared to the change in water. However, the free energy surface of dissociation, reaction pathway as well as the estimated rate constants in these three systems are considerably different.

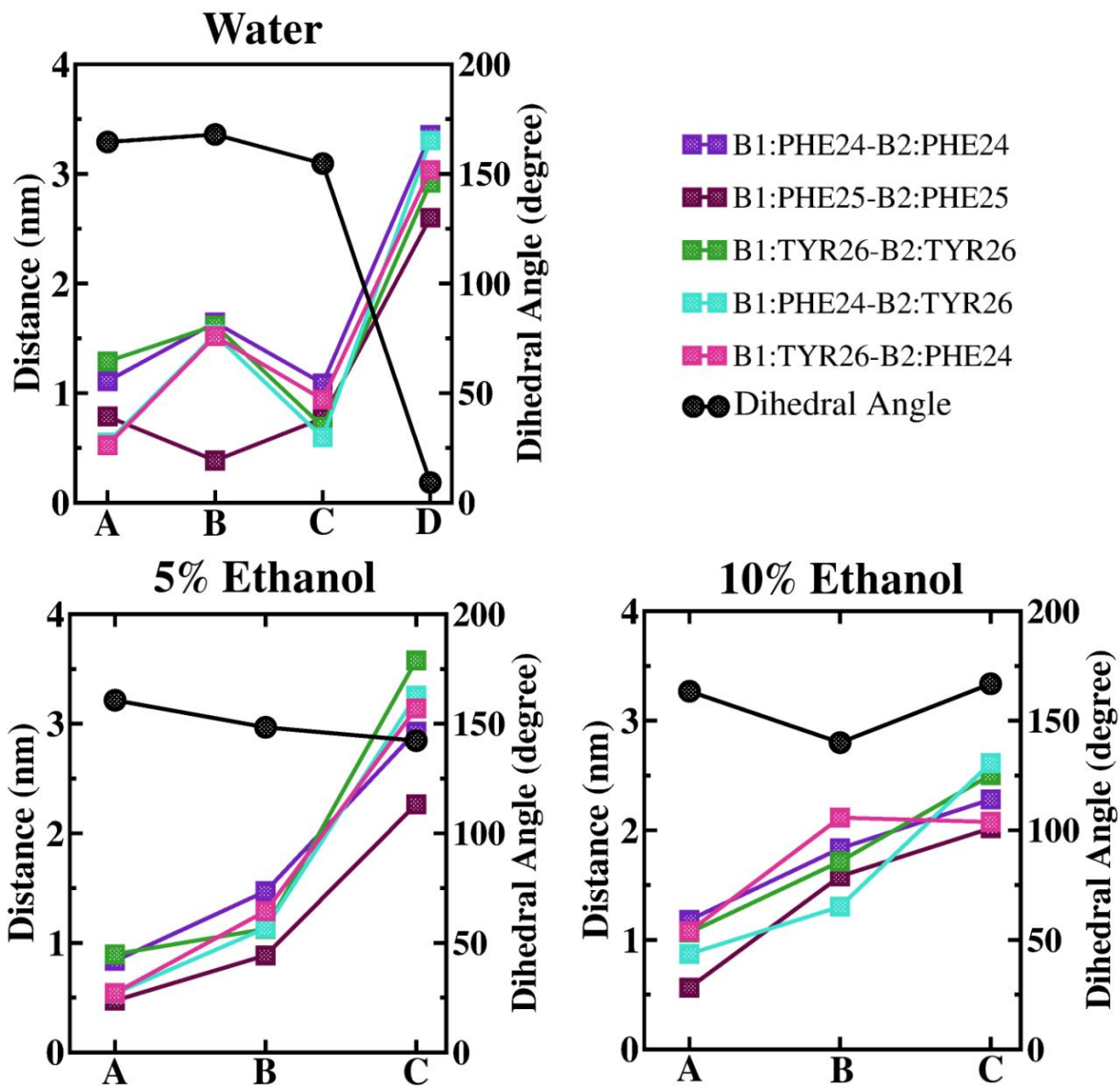

*Figure 8: Change of dihedral angle (monomer1:PHE24-monomer1:TYR26-monomer2:TYR26-monomer2:PHE24) and distances between residues of antiparallel β-sheet in the different stable and intermediate states on the minimum energy pathway of dimer dissociation of insulin.*

To further investigate the effect of ethanol on the dimer dissociation process, we analyze the change in the monomer folding in the dimeric unit in presence of ethanol in the medium. In **Figure 9(a)**, we have plotted a cross distance map of all the residues centers of masses of two monomers forming the dimer in state 'A' in water (**Figure 9(a)**). Here, purple colored region signifies the distant residues and blue colored region highlights the close contacts between two monomers. Not surprisingly, C-terminals of the B-chains (residues 43-48) of two monomers are in closest contacts with each other. This region forms the antiparallel β-sheet between two monomers that provides stability to the dimeric form. On the other hand, the N-termini regions of the B-chains (residues 22-26) of two monomeric units remain farthest from each other.

We have computed the cross distance map of two monomeric units corresponds to the most stable state 'A' in both 5% and 10% water-ethanol mixture and report the difference map between the two distance maps, that is, in water and in 5%/10% water-ethanol mixture in **Figure 9(b-c)**. The difference maps clearly show some residues of two monomers move away (purple region) and some other residues come closer (blue region) compared to the stable dimeric structure in water. We observe these following most prominent changes in the difference map of 5% ethanol-water mixture:

(i) A-chain residue TYR20 of monomer 1 and some of its neighboring residues move away from B-chain residue PHE25 of monomer 2 and its neighbors.

(ii) A-chain residue TYR19 of monomer 1 moves apart from the same residue and its surroundings of monomer 2.

(iii) A-chain residues ILE2, VAL3 of monomer 1 increase their distance from C terminal region of A-chain of monomer 2 (containing TYR19, ASN21).

(iv) B-chain residues PHE24-TYR26 of monomer 1 move away from the B-chain residues PHE24-TYR26 of monomer 2(β-sheet region).

(v) B-chain residue, PHE1 of both monomers comes close to N-terminal region of B-chain of the other monomer.

(vi) PHE1 of monomer 1 (B-chain) comes close to A-chain residues of monomer 2 close to ILE10.

These changes suggest different solvation of hydrophobic and polar residues by ethanol molecules in the solvent medium that modifies the conformation of both monomers significantly. In case of 10% ethanolic solution, mostly all the residues of two monomers in contacts with each other in pure water move apart.

The signature for the antiparallel β sheet is modified in both the cases, however, the change in the folding of overall protein structure is more pronounced for both 5% and 10% water-ethanol mixtures. These results claim that the stabilizing antiparallel β sheet does not get affected much along the minimum energy pathway upto it reaches to very low value of number of contacts. However, the overall protein conformation experiences a dramatic change in the presence of ethanol. Here we have shown the change in protein folding for the most stable dimeric state (state 'A') in different medium, however, in the other intermediate states the changes become more pronounced that drives the reaction in different pathway with different energy barrier and reaction rate in the presence of ethanol.

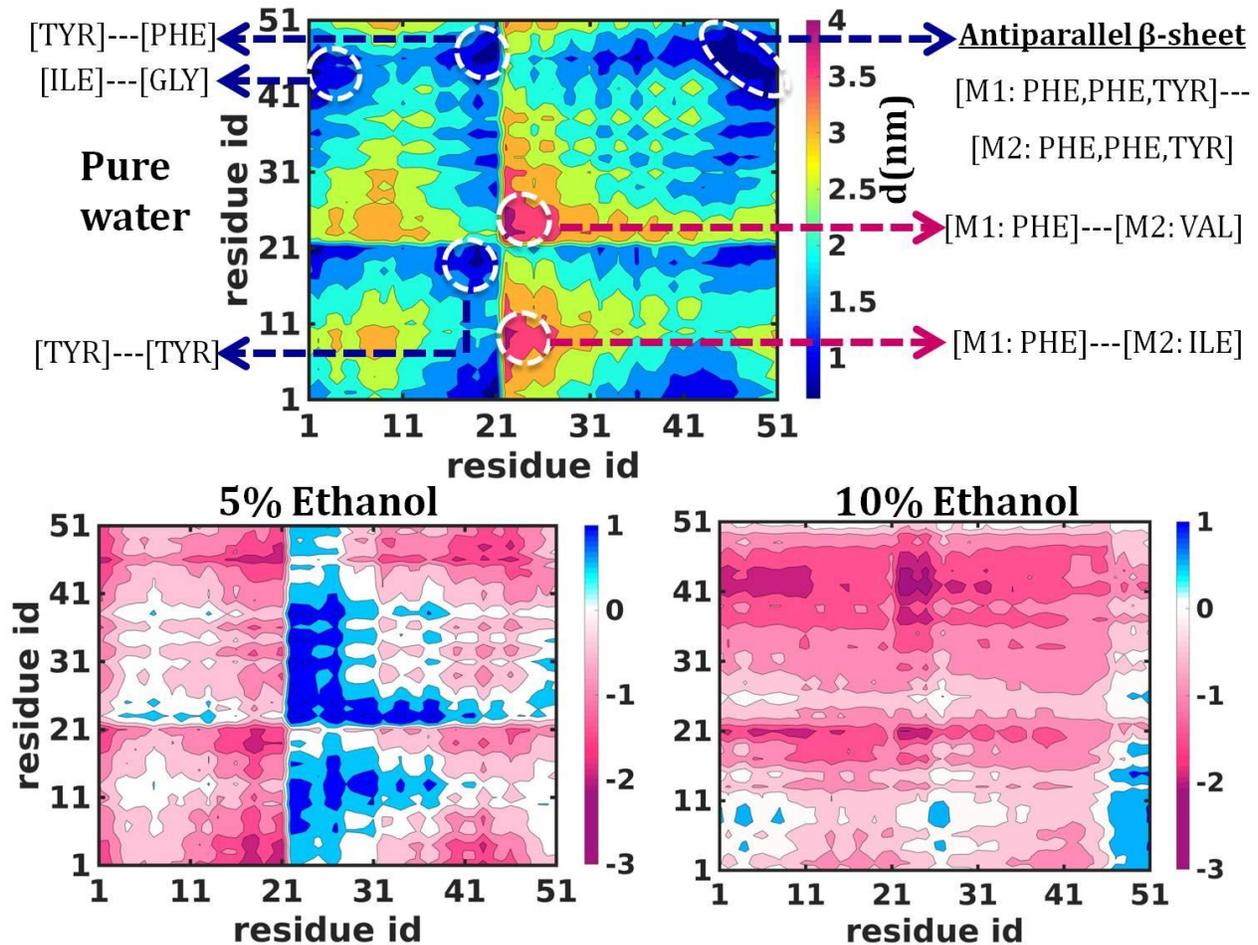

*Figure 9:(a) Distance map of stable insulin dimer (conformer 'A') in water. x and y axis contains the residue id of monomer 1 and monomer 2 respectively. Dark blue regions indicate the residues in dimer that are close to each other and red regions are for the farthest residue pairs. The distance scale is in nm. Difference map of distance in water with respect to (b) 5% water-ethanol mixture, (c) 10% water-ethanol mixture ($d_{water}-d_{Ethanol5\%/10\%}$). Blue regions point out the residues that come close to each other in presence of ethanol and red regions is for the residues that move away from each other.*

# D. Enhancement of dissociation rate by preferential solvation in ethanol

In all the previous sections, we have discussed about the change in free energy of the intermediate states along the minimum energy pathway of dimer dissociation in water and water-ethanol binary mixtures. Subsequently, we have discussed the structural changes of antiparallel β-sheet as well as the whole protein conformation in water and different ethanol concentration in the medium. During the dimer dissociation when two monomers move apart from each other, bulk water molecules come into their junction at the cost of entropy[49], however, the extra solvation energy of the monomers in the dimer forming surface contributes to the enthalpic stabilization. Moreover, the protein conformations also get modified along the reaction pathway which contributes to a huge entropic change and determines stability of intermediate state and hence the reaction pathway. In case of water-ethanol mixture, the nature of solvation is different as ethanol is an amphiphilic solvent and can have a significant interactions with the hydrophobic residues of the protein molecule and also with the polar residues. Hence, in that case the different mode of entropic change with different protein conformation decides the reaction pathway as well as change in free energy.

This is apparent from FES that where some states are shallow and wide giving rise to entropy, in some other cases minima is deep and narrow. Note that the dimer has substantially low free energy than the widely separated monomers. This stabilization is not only by hydrophobic interaction and hydrogen bond but also due to entropic stabilization of bulk solvent molecules.

Now, we want to analyze the difference of solvation of the antiparallel β-sheet (that mainly stabilizes the insulin dimer) by ethanol molecules in two water-ethanol mixtures. We have shown two snapshots of the dimer junction in state 'A' and state 'C' (along the MEP) solvated with ethanol molecules in the 10% water-ethanol mixtures in **Figure 10(a-b)**. As we stated earlier, more number of ethanol molecules comes into the junction of two monomers in the state 'C' loosing their entropy, however, interaction of ethanol molecules with the exposed residues gives enthalpic stabilization. In the two snapshots, clearly the protein conformation modifies significantly that also changes the entropy of the system.

We have shown the Radial distribution function of ethanol molecules around the residues of PHE24, PHE25 and TYR26 forming the antiparallel β-sheet of the dimer in **Figure 10(c-e)** for 5% water-ethanol mixture and **Figure 10(f-h)** for 10% water-ethanol mixture for all the intermediate states during dimer dissociation. We should note here is that these three residues in two individual monomers in the dimeric form exhibits quite different distribution of ethanol molecules around it. Here we have shown the RDF around the three residues of monomer 1, in ***Supporting information (Figure S-1),*** other RDF plots for monomer 2 are shown. For most of the residues, except PHE25 in 5% water-ethanol mixture, in state 'A' they experience least number of ethanol molecules distributed around them as expected and along the MEP the distribution changes significantly. Surprizingly, in state 'B' of 5% ethanol concentration, TYR26 of monomer 1 experiences considerable solvation by ethanol.

In **Figure 10(i,j)**, we have plotted RDF of ethanol molecules around the said three residues in a isolated monomeric form of insulin in 5% and 10% water-ethanol mixtures. The most important thing to note here is, in state 'C' for both 5% and 10% mixtures, RDFs of all residues in partly separated dimeric structure are considerably different from the RDFs of isolated monomer in corresponding water-ethanol mixture for both the monomers (**Figure 10** and **Figure S-1** of ***Supporting information)***. However, in pure water, the RDFs of water around the residues in state 'D' of the dimer are shown to be similar to that of monomeric insulin (**Figure S-2** of *Supporting information*).

Now, if we correlate these results of solvation with the microscopic structural changes in the intermediate states discussed in section III.C, the modification in the free energy surface that increases the rate of dissociation in presence of ethanol can be explained. **Figure 3** and **Figure 4** show the FES of insulin dimer dissociation in 5% and 10% water-ethanol mixture. In both the cases, we see the intermediate state 'C' corresponds to very low number of contacts but considerably close association with $R_{MM}$~3 nm. Unlike the FES in water, these minima are significantly stable in both water-ethanol mixtures, even in 10% water-ethanol mixture this is the global minima of the FES. The stability of these partially separated dimeric state basically increases both the dissociation and association rate of insulin dimer in presence of ethanol in the medium.

In contrast to the insulin dimer dissociation in water, we have shown in section III.C that the relative orientation of antiparallel β-sheet structure(connected by backbone hydrogen bonding) is still maintained in the state 'C' for both the water-ethanol mixtures though the distance between amino acid residues increases significantly and number of contacts become almost zero (**Figure 8**). The preferential solvation at the dimer interface causes this long range interaction between the antiparallel β-sheet forming regions of the two monomers maintaining the relative orientation of the backbones that stabilizes the intermediate state 'C'.

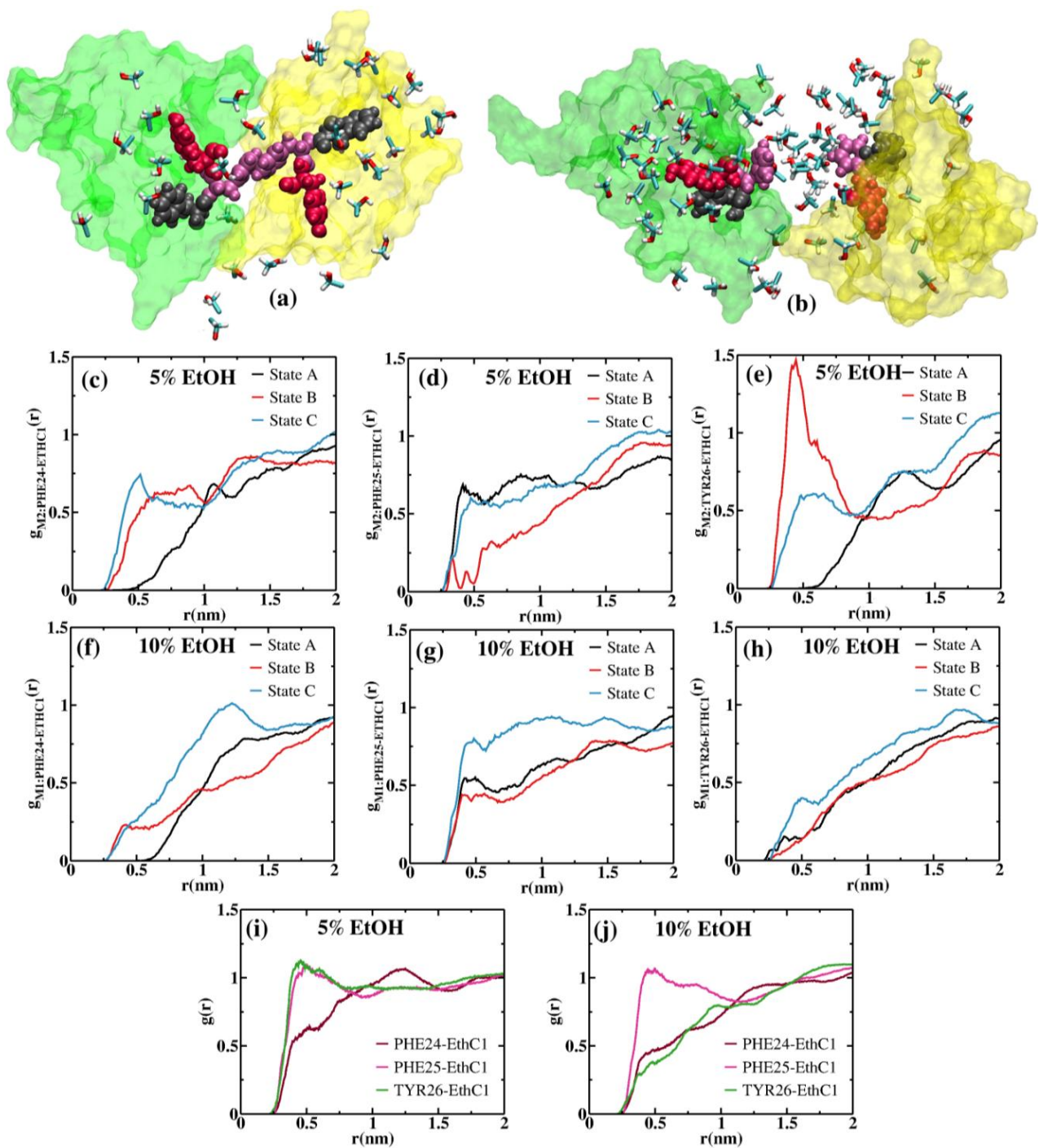

*Figure 10: (a) Snapshots of ethanol distribution in the junction of insulin dimer in 10% water ethanol binary mixture for (a) state 'A' and (b) state 'C'. Radial distribution functions of ethanol molecules around PHE24, PHE25, TYR26 of monomer-1 forming the antiparallel β-*

*sheet in different intermediate states along the minimum energy path: (c-e) RDFs in 5% water-ethanol mixture, (f-h) RDFs in 10% water-ethanol mixture. RDFs of the ethanol molecules around PHE24, PHE25, TYR26 of isolated insulin monomer in (i) 5% water-ethanol and (j) 10% water ethanol mixtures.*

## IV. Conclusion

Earlier several studies have addressed the problem of association of two protein monomers forming the dimer. In the association problem, mutual diffusion of the two monomers to reach the distance of close contact and the subsequent rotational alignment could play the rate determining processes. Bakker et al. have discussed these issues in detail and argued that in association of two monomers, the two diffusive processes mentioned above might predominantly determine the rate of association. These authors further pointed out that in such a scenario, the later and the last barrier crossing step might play only a minor role [25].

We have observed here that the scenario is almost completely opposite in the case of protein dissociation. Here the initial process of spatial separation could be the rate determining step. Another difference between the study of association and dissociation is that the former is a *'many to one'* mapping scenario whereas the latter is a case of *'one to many'* mapping process.

Ethanol is unique among amphiphilic solvents because it is completely miscible with water at all compositions. The mixture displays a well-known eutectic point close to pure ethanol composition, at 94% EtOH, and displays an unusually low freezing temperature of $-20^0$ C. Physico-chemical properties of water-ethanol mixtures have been studied in great detail recently, including studies focussed to conformational change of folded proteins in their native state. High level experiments including 2D-IR studies have been carried out to understand the nature of its stability and also the change due to the presence of ethanol in water-ethanol binary mixture. These experiments find a reduced stability of the insulin dimer due to the presence of EtOH in the solvent.

From a biological point of view, ethanol is of considerable importance as it is a component of many common drinks. Sometimes, concentration of ethanol in body can increase substantially for a short duration. Even though biochemical processes are slow, the increased concentration may stay long enough to have an effect. Historically, small amounts of alcohol have been advocated to be good for human health while excess could have adverse effects. However, nothing is known conclusively.

From a chemical point of view, ethanol bears a strong amphiphilic character, with the ethyl group sufficiently large in size to have hydrophobic interaction with such groups as phenyl alanine, isoleucine. Therefore, it can serve a denaturant. Theoretical studies have shown that the denaturant effect is rather weak, in comparison to for example that of dimethyl sulfoxide. However, as the present study shows ethanol could have a serious role to play in biochemical processes.

The study of stability of insulin dimer is itself interesting and deserves attention. Insulin dimer seems to have an ephemeral character. Its main biological function, as far is known, is to serve as an intermediary between the biologically active monomer and the storage house that is the hexamer. The dimer is held together both by four hydrogen bonds between the monomer groups and also hydrophobic interactions. Compared to two monomers, it is stabilized by a free energy of 51.9 kJ/mole.

In presence of ethanol, both the hydrophobic interactions and the hydrogen bonds of the dimer come under pressure because ethanol can play both these roles and solvate individual monomers to an extent that the dimer becomes less stable compared to that in water. The dimer of course does not lose its full stability but becomes less stable. Our body uses up the storehouse of insulin according to the requirement of monomeric insulin in our bloodstream in order to control blood glucose level. Because of the reduction in the dissociation free energy barrier, the dimer can produce monomers in presence of a small amount of ethanol even if our body does not require it at that moment. This might lead to loss of effective insulin molecules which in turn result in hypoinsulinemia.

Our present study here shows that in addition to causing a reduction in stability, the dissociation reaction pathway undergoes significant change due to the presence of ethanol molecules in the water-ethanol binary mixture, even at a relatively low concentration of 10%. The difference in reaction free energy surface (R-FES) between pure water and 10% water-ethanol binary mixture is quite striking.

The computed FES in pure water exhibits pronounced ruggedness due to the presence of a number of quasi-stable intermediate states. In this work we have been able to identify the intermediate states. In the initial phase of dimer dissociation in water, the contacts between amino acid residues from the different monomers reduces to a great extent while maintaining a similar inter-monomeric distance. The reaction pathway is characterized by a broad and high transition state, with a barrier energy ~50 kJ/mol.

Significant changes to this scenario occurs in the presence of ethanol. The presence of the latter smoothens out the ruggedness. The barrier height decreases and the free energies of the intermediate states also decrease. We have extracted the most probable conformation of the intermediate states on the minimum energy pathway and analyzed the microscopic structural changes of protein in the presence of ethanol. We find that in the presence of ethanol, this reaction pathway changes considerably due to the entropy-enthalpic competition and different stabilisation of the intermediate states. *The distribution of water and ethanol around the dimer forming region reveals that in presence of ethanol, the two antiparallel β- sheet forming region can experience a long range interaction and stabilize the intermediate state 'C' where the number of contacts is close to zero but the inter-monomeric distance is not too small (~3 nm). This quasi-stable intermediate in turn reduces the barrier height and enhances the rate of insulin dimer dissociation, explaining the role of ethanol.*

We also quantify the decrease in the barrier height of dissociation reaction. Furthermore, we obtain estimates of the rate of reaction of dimer dissociation for all the three systems. The results agree well with the available experimental results for the insulin dimer dissociation in water. The recent experimental results and the present theoretical analysis suggest that insulin dimer dissociation can serve as a hugely important system to study protein dissociation. This system while contains the complexity intrinsic to proteins, yet at the same time sufficiently simple to allow detailed theoretical and experimental investigations of the system.

We believe that the present study has captured many of the essential aspects of a protein dimer dissociation reaction that can be of use in future research.

## Conflicts of interest

The authors declare no conflicts of interest.

## Acknowledgement:

The work was supported partly by the Department of Science and Technology (DST), Govt. of India, Sir J. C. Bose fellowship, the Council of Scientific and Industrial Research (CSIR), India and the University Grants Commission (UGC), India.